\documentclass[runningheads]{llncs}
\usepackage[T1]{fontenc}
\usepackage{graphicx}
\usepackage{amsmath}
\usepackage{amssymb}
\usepackage{mathtools}

\DeclarePairedDelimiter\floor{\lfloor}{\rfloor}

\makeatletter
\newcommand{\printfnsymbol}[1]{%
  \textsuperscript{\@fnsymbol{#1}}%
}
\makeatother

\makeatletter
\def\@fnsymbol#1{\ensuremath{\ifcase#1\or *\or \dagger\or \ddagger\or
   \mathsection\or \mathparagraph\or \|\or **\or \dagger\dagger
   \or \ddagger\ddagger \else\@ctrerr\fi}}
    \makeatother

\newenvironment{alphafootnotes}
  {\par\edef\savedfootnotenumber{\number\value{footnote}}
   
   \setcounter{footnote}{0}}
{\par\setcounter{footnote}{\savedfootnotenumber}}

\begin{document}
\title{Simulation of Arbitrary Level Contrast Dose in MRI Using an Iterative Global Transformer Model}
\author{Dayang Wang\inst{1,2}\thanks{The first two authors are equal contributors and co-first authors.}, Srivathsa Pasumarthi\inst{1}\printfnsymbol{1}\thanks{Corresponding author}, Greg Zaharchuk\inst{3}, Ryan Chamberlain\inst{1}}

\authorrunning{D. Wang, S. Pasumarthi, et al.}

\institute{
Subtle Medical Inc., Menlo Park CA 94025, USA\\
\email{srivathsa@subtlemedical.com}
\and
University of Massachusetts Lowell, Lowell MA 01854, USA
\and
Stanford University, Stanford CA 94035, USA\\
}
\maketitle
\begin{alphafootnotes}
\begin{abstract}
\footnote{\textbf{Paper accepted in the 26th International Conference on Medical Image Computing and Computer Assisted Interventation (MICCAI 2023) to be held in Vancouver, Canada from October 8-12, 2023.}}
Deep learning (DL) based contrast dose reduction and elimination in MRI imaging is gaining traction, given the detrimental effects of Gadolinium-based Contrast Agents (GBCAs). These DL algorithms are however limited by the availability of high quality low dose datasets. Additionally, different types of GBCAs and pathologies require different dose levels for the DL algorithms to work reliably. In this work, we formulate a novel transformer (Gformer) based iterative modelling approach for the synthesis of images with arbitrary contrast enhancement that corresponds to different dose levels. The proposed Gformer incorporates a sub-sampling based attention mechanism and a rotational shift module that captures the various contrast related features. Quantitative evaluation indicates that the proposed model performs better than other state-of-the-art methods. We further perform quantitative evaluation on downstream tasks such as dose reduction and tumor segmentation to demonstrate the clinical utility.

\\

\end{abstract}
\section{Introduction}
Gadolinium-Based Contrast Agents (GBCAs) are widely used in MRI scans owing to their capability of improving the border delineation and internal morphology of different pathologies and have extensive clinical applications\cite{minton2021contrast}. However, GBCAs have several disadvantages like contraindications in patients with reduced renal function\cite{grobner2006gadolinium}, patient inconvenience, high operation costs and environmental side effects\cite{brunjes2020anthropogenic}. Therefore, there is an increased emphasis on the paradigm of \textit{"as low as reasonably achievable"} (ALARA)\cite{uffmann2009digital}. To tackle these concerns of GBCAs, several dose reduction\cite{pasumarthi2021generic,gong2018deep} and elimination approaches\cite{kleesiek2019can} have been proposed. However, these deep learning(DL)-based dose reduction approaches require high quality low-dose contrast-enhanced (CE) images paired with pre-contrast and full-dose CE images. Acquiring such a dataset requires modification of the standard imaging protocol and involves additional training of the MR technicians. Therefore, it is important to simulate the process of T1w low-dose image acquisition, using images from the standard protocol. Moreover, it is crucial for these dose reduction approaches to establish the minimum dose level required for different pathologies as these are dependent on the scanning protocol and the GBCA compound injected. Therefore the simulation tool should also have the ability to synthesize images with multiple contrast enhancement levels, that correspond to multiple arbitrary dose levels.

Currently MRI dose simulation is done using physics-based models\cite{sourbron2011tracer}. However, these physics-based methods are dependent on the protocol parameters and the type of GBCA and their relaxation parameters. Deep learning (DL) models have been widely used in medical imaging application due to their high capacity, generazibility, and transferability\cite{wang2021ted,liu2022one}. The performance of these DL models heavily depend on the availability of high quality data. There is a dearth of data-driven approaches to MRI dose-simulation given the lack of diverse ground truth data of the different dose levels. To this effect, we introduce a vision transformer based DL model\footnote{A part of this work was presented as a poster in the conference of International Society for Magnetic Resonance in Medicine (ISMRM) 2023, held in Toronto.} that can synthesize brain\footnote{Refer Supplementary Material Figure 1(b) for examples on Spine MRI.} MRI images that correspond to arbitrary dose levels, by training on a highly imbalanced dataset with only T1w pre-contrast, T1w 10\% low-dose, and T1w CE standard dose images. The model backbone consists of a novel Global transformer (Gformer) with subsampling attention that can learn long-range dependencies of contrast uptake features. The proposed method also consists of a rotational shift operation that can further capture the shape irregularity of the contrast uptake regions. We performed extensive quantitative evaluation in comparison to other state-of-the art methods. Additionally, we show the clinical utility of the simulated T1w low-dose images using downstream tasks. To the best of our knowledge, this is the first DL based MRI dose simulation approach.

\section{Methods}
\textbf{Iterative learning design:} DL based models tend to perform poorly when the training data is highly imbalanced \cite{antipov2017face}. Furthermore, the problem of arbitrary dose simulation requires the interpolation of intermediate dose-levels using a minimum number of data points. Iterative models \cite{liu2019ifr,shan2019competitive} are suitable for such applications as they work on the terminal images to generate step-wise intermediate solutions. We first utilize this design paradigm for the dose simulation task and train an end-to-end model on a highly imbalanced dataset where only T1w pre-contrast, T1w low-dose, and T1w post-contrast are available.
\begin{figure}
\centering
\includegraphics[width=0.8\textwidth]{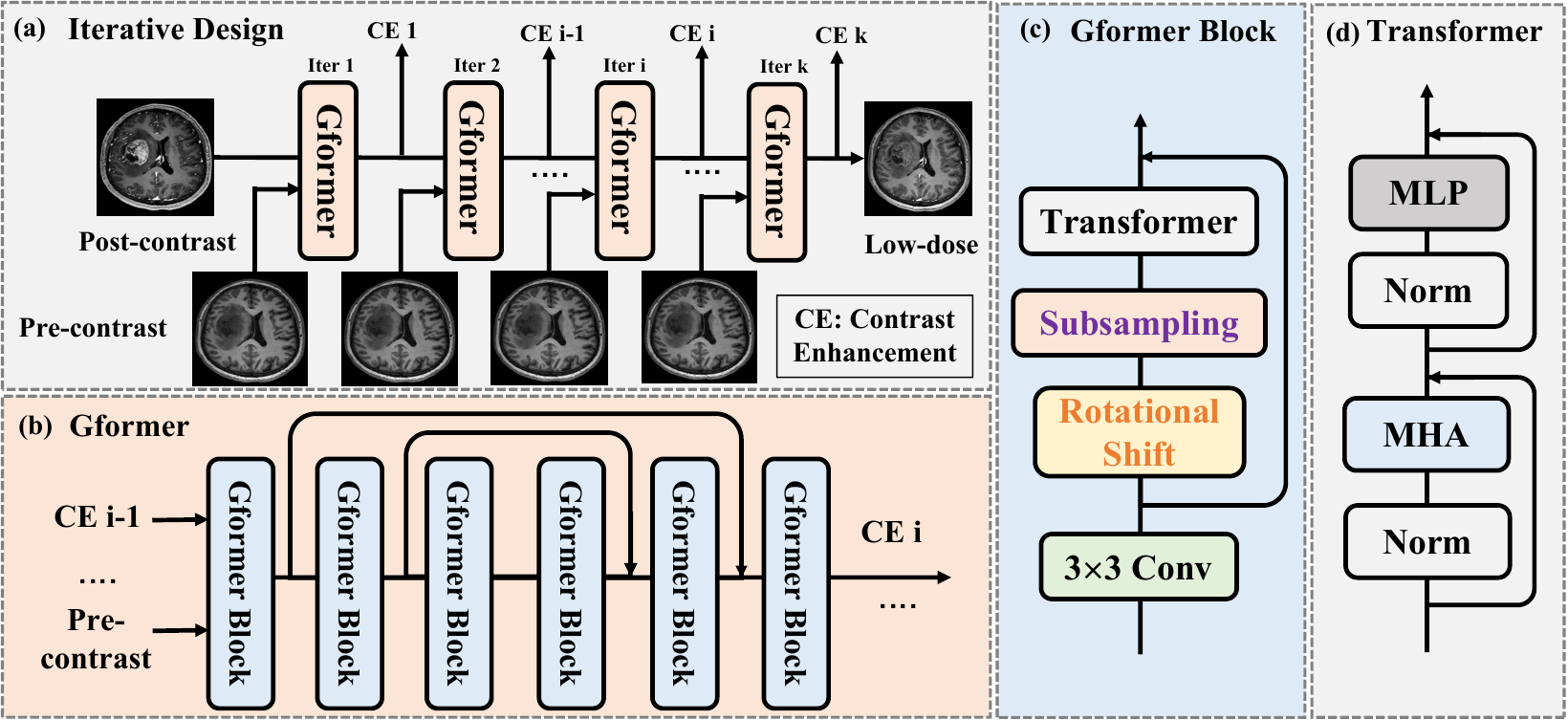}
\caption{(a) The proposed model based on iterative learning design. (b) The overall architecture of the proposed Gformer as the backbone network. (c) Layers inside the Gformer block. (d) Illustration of a typical vision transformer block.}
\label{wholeflow}
\end{figure}

As shown in Fig. 1(a), the proposed iterative model $\boldsymbol{\mathcal{G}} = \boldsymbol{\mathcal{F}} \circ \boldsymbol{\mathcal{F}} \circ \cdots \circ \boldsymbol{\mathcal{F}}$ learns a transformation from the post-contrast to the low-dose image in $k$ iterations, where $\boldsymbol{\mathcal{F}}$ represents the base model.
At each iteration $i$, the higher enhancement image from the previous step and the pre-contrast images are fed into $\boldsymbol{\mathcal{F}}$ to predict the image with lower enhancement. The iterative model can be formulated as follows,
\begin{equation}
\begin{cases}
    \widehat{\mathbf{P_i}} = \boldsymbol{\mathcal{F}} (\widehat{\mathbf{P_{i-1}}},\mathbf{P_{pre}})\\

    \widehat{{\mathbf{P_{low}}}} = \overbrace{\boldsymbol{\mathcal{F}} \circ \boldsymbol{\mathcal{F}} \circ \cdots \circ \boldsymbol{\mathcal{F}} }^{k} (\mathbf{P_{post}},\mathbf{P_{pre}}),

\end{cases}
\end{equation}
where $\mathbf{P_{pre}}$, $\mathbf{P_{post}}$, and $\widehat{{\mathbf{P_{low}}}}$ denote the pre-contrast, post-contrast, and predicted low-dose images, respectively and $\widehat{\mathbf{P_{i-1}}}$ denotes the image with a higher enhancement than $\widehat{\mathbf{P_i}}$. This way, the intermediate outputs $\{\widehat{\mathbf{P_i}}\}^k_{i=1}$ having different enhancement levels, correspond to images with different contrast dose level with a uniform interval. This iterative model essentially learns a gradual dose reduction process, in which each iteration step removes a certain amount of contrast enhancement from the full-dose image.
\\
\\
\noindent\textbf{Loss functions and model convergence:} The proposed iterative model aims to learn a mapping from the post-contrast \& pre-contrast images to the synthesized low-dose images $\widehat{{\mathbf{P_{low}}}}$ and is trained with the true 10\% low-dose image ${\mathbf{P_{low}}}$ as the ground truth. We used the $\mathbf{L_1}$ and structural similarity index measure (SSIM) losses. To tackle the problem of gradient explosion or vanishing, \textit{"soft labels"} are generated using linear scaling. These \textit{"soft labels"} serve as a reference to the intermediate outputs during the iterative training process and also aid model convergence, without which the model has to directly learn from post-contrast to low-dose. Given $k$ iterations, the \textit{soft label} $\{\boldsymbol{\mathcal{S}_i}\}^{k-1}_{i=1}$ for iteration $i$ is calculated as follows:
\begin{equation}
\label{eqn:soft_label}
    \boldsymbol{\mathcal{S}_i} = \mathbf{P_{pre}} +  [\gamma + (1-\gamma) \frac{k-i}{k}] \times \mathrm{ReLU}(\widetilde{\mathbf{P_{post}}} - \widetilde{\mathbf{P_{pre}}} - \tau),
\end{equation}
where $\widetilde{\mathbf{P_{post}}}$ and $\widetilde{\mathbf{P_{pre}}}$ denote the skull-stripped post-contrast and pre-contrast images. $\gamma=0.1$ represents the dose level of the final prediction, and $\tau=0.1$ denotes the threshold to extract the estimated contrast uptake $\boldsymbol{\mathcal{U}} = \mathrm{ReLU}(\widetilde{\mathbf{P_{post}}} - \widetilde{\mathbf{P_{pre}}} - \tau)$. Finally, the total losses are calculated as
\begin{equation}
    \boldsymbol{\mathcal{L}_{\mathrm{total}}} = \alpha \cdot \sum_{i=1}^{k-1}
    \boldsymbol{\mathcal{L}_{\mathrm{e}}}(\widehat{\mathbf{P_i}},\boldsymbol{\mathcal{S}_i})+ \beta \cdot \boldsymbol{\mathcal{L}_{\mathrm{e}}}(\widehat{{\mathbf{P_{low}}}},{\mathbf{P_{low}}}).
\end{equation}
Where $\boldsymbol{\mathcal{L}_{\mathrm{e}}} = \boldsymbol{\mathcal{L}_{\mathrm{L1}}} + \boldsymbol{\mathcal{L}_{\mathrm{SSIM}}}$ and $\alpha=0.1$ and $\beta=1$. The \textit{"soft labels"} are assigned a small loss weight so that they do not overshadow the contribution of the real low-dose image. Additionally, in order to recover the high frequency texture information and to improve the overall perceptual quality, adversarial \cite{goodfellow2020generative} and perceptual losses \cite{johnson2016perceptual} are applied on $(\widehat{{\mathbf{P_{low}}}},{\mathbf{P_{low}}})$ with a weight of $0.1$.
\\
\\
\noindent\textbf{Global transformer (Gformer):}
Transformer models have risen to prominence in a wide range of computer vision applications \cite{liu2022one,liu2021swin}. Traditional Swin transformers compute attention on non-overlapping local window patches. To further exploit the global contrast information, we propose a hybrid global transformer (Gformer) as a backbone for the dose simulation task. As illustrated in Fig. 1(b), the proposed model design includes six sequential Gformer blocks as the backbone module with shortcuts. As shown in Fig. 1(c), the Gformer block contains a convolution block, a rotational shift module, a sub-sampling process, and a typical transformer module. The convolution layer extracts granular local information of the contrast uptake while the self-attention emphasizes more on the coarse global context, thereby paying attention to the overall contrast uptake structure.
\\
\\
\textbf{Subsampling attention:} The sub-sampling is a key element in the Gformer block which generates a number of sub-images from the whole image as attention windows as shown in Fig \ref{rotation}. Gformer performs self-attention on the sub-sampled images, which encompasses global contextual information with minimal self-attention overhead on small feature maps. Given the entire feature map $\mathbf{M_e} \in \mathbb{R}^{b \times c \times h \times w}$, where $b, c, h$, and $w$
are the batch size, channel dimension, height, and width, respectively, the subsampling process aggregates the strided positions to the sub-feature maps as follows,
\begin{equation}
    \{\mathbf{M_{s}}\}^{{d}^2-1}_{s=0} = \{ \mathbf{M}[\ : \ , \ : \ , \ i:h:d \ , \ j:w:d \ ] \}_{i=0,j=0}^{d-1,d-1},
\end{equation}
where $d$ denotes sampling a position every $d$ pixels, and $\mathbf{M_s} \in \mathbb{R}^{b \times c \times \frac{h}{d} \times \frac{h}{d}}$ is the subsampled feature map. We set $h, d=0$ to avoid any information loss during subsampling. These $d^2$ sub-feature maps are stacked onto the batch dimension as the attention windows for the transformer block shown in Fig \ref{wholeflow}(c).

\begin{figure}
\centering
\includegraphics[width=0.8\textwidth]{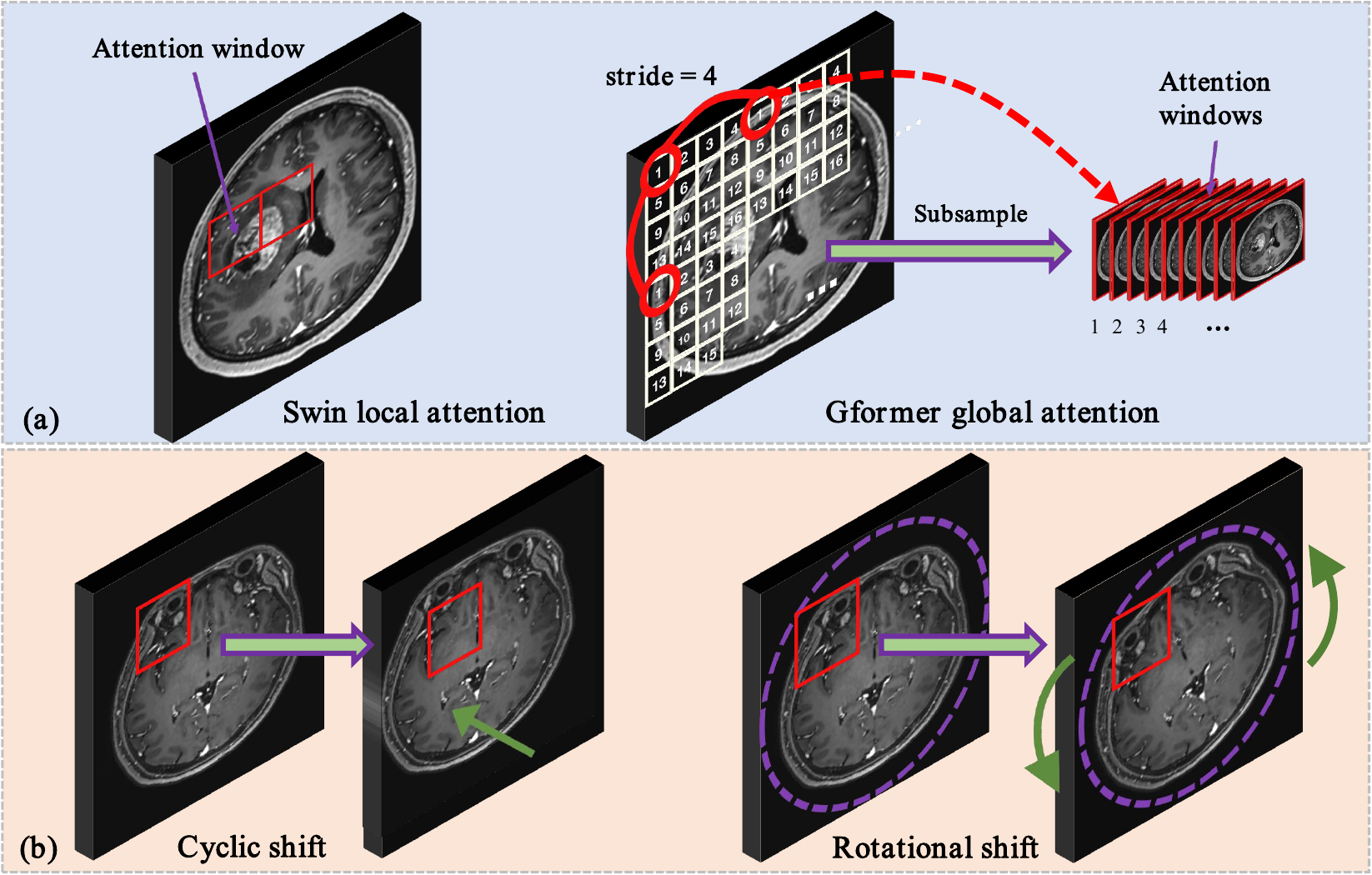}
\caption{(a)Illustration of the sub-sampling process with the stride of 4 as global attention window in Gformer block in comparison to local attention in Swin transformer. All pixels within the same number aggregate to the same sub-image. (b) Depiction of how rotational shift can enhance diverse contextual information fusion across layers compared to cyclic shift in Swin transformer.}
\label{rotation}
\end{figure}

\noindent\textbf{Rotational shift:} Image rotation has been widely used as a data augmentation technique in preprocessing and model training. Here, to further capture the heterogeneous nature of the contrast uptake areas, we employ the rotational shift as a module to facilitate the representation power of the Gformer. To prevent information loss on the edges due to rotation, only small angles (e.g., $10^{\circ}$, $20^{\circ}$) are used for rotation and residual shortcuts are also applied. Specifically, given the feature map $\mathbf{M_o}  \in \mathbb{R}^{b \times c \times h \times w}$, rotational shift is performed around the vertical axis of height/width. The rotated feature map $\mathbf{M_r} \in \mathbb{R}^{b \times c \times h \times w}$ is obtained by the following equation:
\begin{equation}
\begin{bmatrix} \boldsymbol{p'}\\ \boldsymbol{q'}\\ \boldsymbol{x'}\\ \boldsymbol{y'} \end{bmatrix} =
\begin{bmatrix}
1 & \ 0 & 0 & 0\\
0 & \ 1 & 0 & 0 \\
0 & \ 0 & cos \boldsymbol{\lambda} & -sin \boldsymbol{\lambda}  \\
0 & \ 0 & sin \boldsymbol{\lambda} & cos \boldsymbol{\lambda} \\
\end{bmatrix}
\begin{bmatrix} \boldsymbol{p} \\ \boldsymbol{q} \\ \boldsymbol{x}- h//2\\ \boldsymbol{y}- w//2\\ \end{bmatrix} + \begin{bmatrix} 0\\ 0\\ h//2\\ w//2  \end{bmatrix} , \
\end{equation}
\begin{equation}
    \mathbf{M_r}(\boldsymbol{p},\boldsymbol{q},\boldsymbol{x},\boldsymbol{y}) = \left\{\begin{matrix}
\mathbf{M_o}(\boldsymbol{p'},\boldsymbol{q'},\floor{\boldsymbol{x'}},\floor{\boldsymbol{y'}}),
\ \mathrm{if} \ \boldsymbol{x'} \in [0,h) \ \mathrm{and} \ \boldsymbol{y'} \in [0,w) \\
 0, \ \mathrm{otherwise},  \\
\end{matrix}\right.
\end{equation}
where $\boldsymbol{\lambda}$ is the rotation angle. $(\boldsymbol{p,q,x,y})$ and $(\boldsymbol{p',q',x',y'})$ denote the pixel index in the feature map tensor before and after rotational shift, respectively.

\section{Experiments and Results}
\textbf{Dataset:}
With IRB approval and informed consent, we retrospectively used 126 clinical cases (113 training, 13 testing) from a internal private dataset\footnote{The dataset used in this study was obtained from Site A and B and are not available under a data sharing license} using Gadoterate meglumine contrast agent (Site A). For downstream task assessment we used 159 patient studies from another site (Site B) using Gadobenate dimeglumine. The detailed cohort description is given in Table \ref{cohort}. The clinical indications for both sites included suspected tumor, post-op tumor follow-up and routine brain. For each patient, 3D T1w MPRAGE
scans were acquired for the pre-contrast, low-dose, and post-contrast images. These paired images were then mean normalized and affine co-registered (pre-contrast as the fixed image) using SimpleElastix \cite{marstal2016simpleelastix}. The images were also skull-stripped, to account for differences in fat suppression, using the HD-BET brain extraction tool \cite{schell2019automated} for generating the \textit{"soft labels"}.

\begin{table}[h]
\caption{Dataset cohort description}
\label{cohort}
\centering
\scalebox{0.8}{
\begin{tabular}{|l|l|l|l|l|l|l|l|l|}
\hline
\textbf{Site} & \textbf{\begin{tabular}[c]{@{}l@{}}Total \\ Cases\end{tabular}} & \textbf{Gender} & \textbf{Age} & \textbf{Scanner} & \textbf{\begin{tabular}[c]{@{}l@{}}Field \\ Strength\end{tabular}} & \textbf{\begin{tabular}[c]{@{}l@{}}TE \\ (sec)\end{tabular}} & \textbf{\begin{tabular}[c]{@{}l@{}}TR \\ (sec)\end{tabular}} & \textbf{\begin{tabular}[c]{@{}l@{}}Flip \\ Angle\end{tabular}} \\ \hline
Site A & 126 & \begin{tabular}[c]{@{}l@{}}55 Females \\ 71 Males\end{tabular} & 48 $\pm$ 16 & \begin{tabular}[c]{@{}l@{}}Philips \\ Insignia\end{tabular} & 3T & 2.97-3.11 & 6.41-6.70 & 8$^\circ$ \\ \hline
Site B & 159 & \begin{tabular}[c]{@{}l@{}}78 Females\\ 81 Males\end{tabular} & 52 $\pm$ 17 & \begin{tabular}[c]{@{}l@{}}GE \\ Discovery\end{tabular} & 3T & 2.99-5.17 & 7.73-12.25 & 8-20$^\circ$ \\ \hline
\end{tabular}}
\end{table}

\textbf{Implementation details:}
All experiments were conducted with a single Tesla V100-SXM2 32GB GPU on a Intel(R) Xeon(R) CPU E5-2698 v4. The Subsampling stride for the six levels of Gformer block were \{4,8,16,16,8,4\}. The Rotational shift angles were \{0,10,20,20,10,0\} across all blocks. The model was optimized using the Adam optimizer with an initial learning rate of 1e-5 and a batch size of 4.

\begin{figure}
\centering
\includegraphics[width=.8\textwidth]{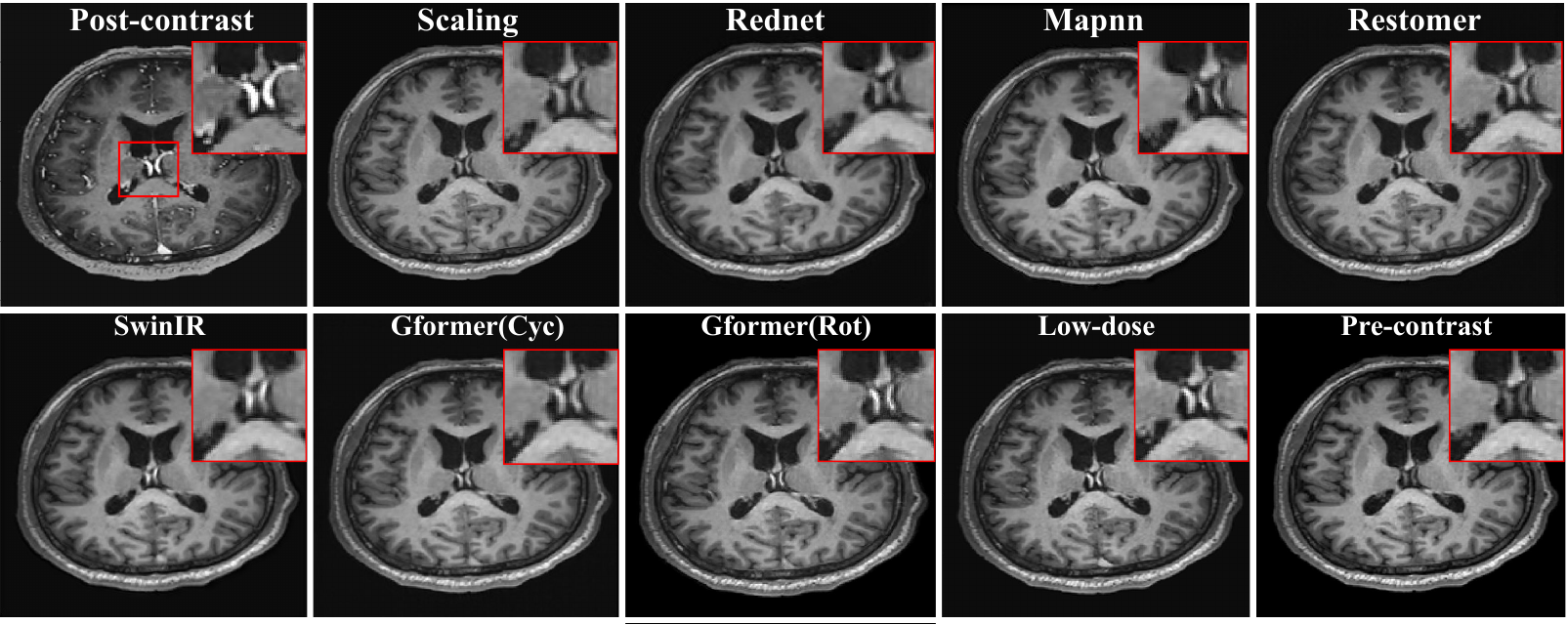}
\caption{The results of the synthesized 10\% dose images from different methods. 'Rot' denotes rotational shift and 'Cyc' indicates cyclic shift. }
\label{compare}
\end{figure}

\noindent\textbf{Evaluation settings:}
We quantitatively evaluated the proposed model using PSNR, SSIM, RMSE, and LPIPS perceptual metrics \cite{zhang2018unreasonable}, between the synthesized and true low-dose images. We replaced the Gformer backbone with other state-of-the-art methods to compare the efficacy of the different methods. Particularly, the following backbone networks were studied: simple linear scaling (\textit{"Scaling"}) approach, Rednet \cite{chen2017low}, Mapnn \cite{shan2019competitive},  Restormer \cite{Zamir2021Restormer}, and SwinIR \cite{liang2021swinir}. Unet \cite{ronneberger2015u} and Swin-Unet \cite{cao2023swin} models were not assessed due to their tendency to synthesize blurry artifacts in the iterative modelling. \textit{throughput} metric (number of images generated per second) was also calculated to assess the inference efficiency.
\\
\\
\noindent\textbf{Evaluation results:}
Fig. \ref{result1}(a) shows that the proposed model is able to generate images that correspond to different dose levels. As shown in the zoomed inset, the hyperintensity of the contrast uptake in these images gradually reduces at each iteration. Fig. \ref{result1}(b) shows that the pathological structure in the synthesized low-dose image is similar to that of the ground truth. Fig. \ref{result1}(c) also shows that the model is robust to hyperintensities that are not related to contrast uptake. Fig. \ref{compare} and Table \ref{quant} show that proposed model can synthesize enhancement patterns that look close to the true low-dose and that it performs better than the other competing methods with a reasonable inference throughput.

\begin{table}[h]
\caption{Quantitative evaluation results of different base methods on the test cases. Bold-faced numbers indicate the best results. }
    \centering
    \scalebox{0.8}{
    \begin{tabular}{c|c|c|c|c|c}
    \hline
    \hline
    Method   &  Throughput &  PSNR (dB)$\uparrow$ & SSIM$\uparrow$   &   RMSE$\downarrow$ &   LPIPS$\downarrow$  \\
    \hline
    Post     &  - & 33.93 $\pm$ 2.88 & 0.93 $\pm$ 0.03 &  0.34 $\pm$ 0.13 &  0.055 $\pm$ 0.016 \\
    Scaling  &  \textbf{0.79} Im/s & 38.41 $\pm$ 2.22 & 0.94 $\pm$ 0.19 &  0.20 $\pm$ 0.05 &  0.027 $\pm$ 0.015\\
    Rednet   &  0.71 Im/s & 40.07 $\pm$ 2.72 & 0.97 $\pm$ 0.01 &  0.17 $\pm$ 0.05 &   0.029 $\pm$ 0.009 \\
    Mapnn    &  0.71 Im/s & 40.56 $\pm$ 1.64  & 0.96 $\pm$ 0.01 & 0.16 $\pm$ 0.05 & 0.023 $\pm$ 0.012 \\
    Restormer & 0.65 Im/s & 40.04 $\pm$ 2.27 & 0.95 $\pm$ 0.01 &  0.16 $\pm$ 0.16 & 0.038 $\pm$ 0.016\\
    SwinIR    & 0.58 Im/s & 40.93 $\pm$ 2.25 & 0.96 $\pm$ 0.01 &  0.15 $\pm$ 0.06 &  0.028 $\pm$ 0.015\\
    \hline
    Gformer*(Cyc)  & 0.69 Im/s & 41.46 $\pm$ 2.14 & 0.97 $\pm$ 0.02 &  0.14 $\pm$ 0.04 &  0.021 $\pm$ 0.007 \\
    Gformer*(Rot)  & 0.65 Im/s & \textbf{42.29 $\pm$ 0.02}  & \textbf{0.98 $\pm$ 0.01} & \textbf{0.13 $\pm$ 0.03} & \textbf{0.017 $\pm$ 0.005}\\
    \hline
    \end{tabular}}
   \label{quant}
    \hfill
\end{table}

\begin{figure}
\centering
\includegraphics[width=.8\textwidth]{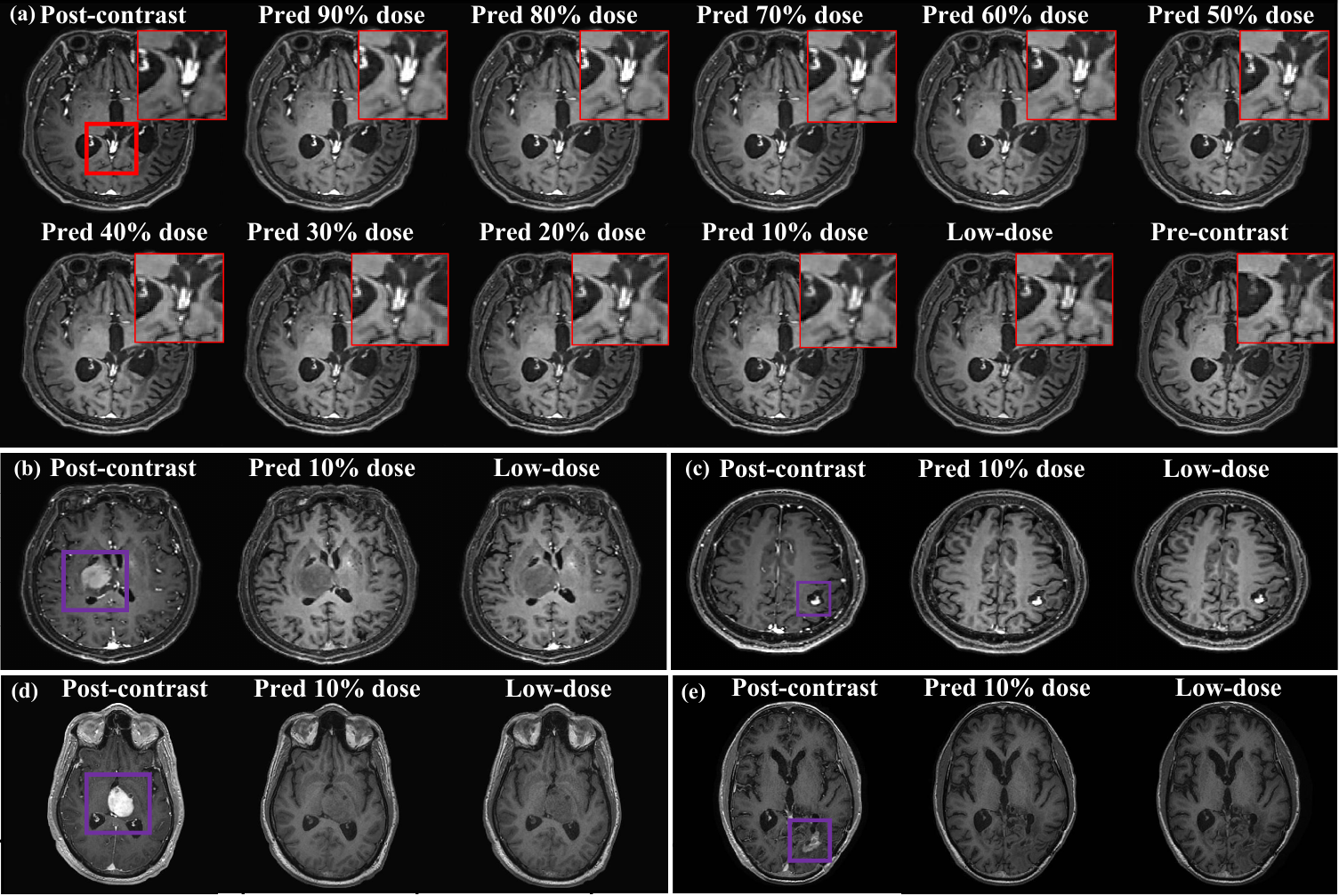}
\caption{(a) Model results showing images with different contrast enhancement corresponding to different dose levels along with the synthesized and true low-dose and pre-contrast. (b)-(c) Two representative slices of the synthesized 10\% dose images. (d)-(e) Two representative slices using a different GBCA.}
\label{result1}
\end{figure}

\noindent\textbf{Quantitative assessment of contrast uptake:}
The above pixel-based metrics do not specifically focus on the contrast uptake region. In order to assess the contrast uptake patterns of the intermediate images, we used the following metrics as described in \cite{bone2022dose}: contrast to noise ratio(CNR), contrast to background ratio(CBR), and contrast enhancement percentage(CEP). The ROI for the contrast uptake was computed as the binary mask of the corresponding \textit{"soft labels"} in Eqn. \ref{eqn:soft_label}. As shown in Fig \ref{contrast}, the value of the contrast specific metrics increases in a non-linear fashion as the iteration step increases.
\\
\\
\noindent
\textbf{Downstream tasks:}
In order to demonstrate the clinical utility of the synthesized low-dose images, we performed two downstream tasks:

\textbf{1) Low-dose to full-dose synthesis}
Using the DL-based algorithm to predict full-dose image from pre-contrast and low-dose images described in \cite{pasumarthi2021generic}, we synthesized T1CE volumes using true low-dose (\textit{T1CE-real-ldose}) and Gformer (rot) synthesized low-dose (\textit{T1CE-synth-ldose}). We computed the PSNR
and SSIM metrics of T1CE vs T1CE-synth/T1CE vs T1CE-synth-sim which are $29.82 \pm 3.90$ dB/$28.10 \pm 3.20$ dB and $0.908 \pm 0.031$/$0.892 \pm 0.026$ respectively. This shows that the synthesized low-dose images perform similar\footnote{p < 0.0001 (Wilcoxon signed rank test)} to that of the low-dose image in the dose reduction task. For this analysis we used the data from Site B.

\begin{figure}
\centering
\includegraphics[width=.7\textwidth]{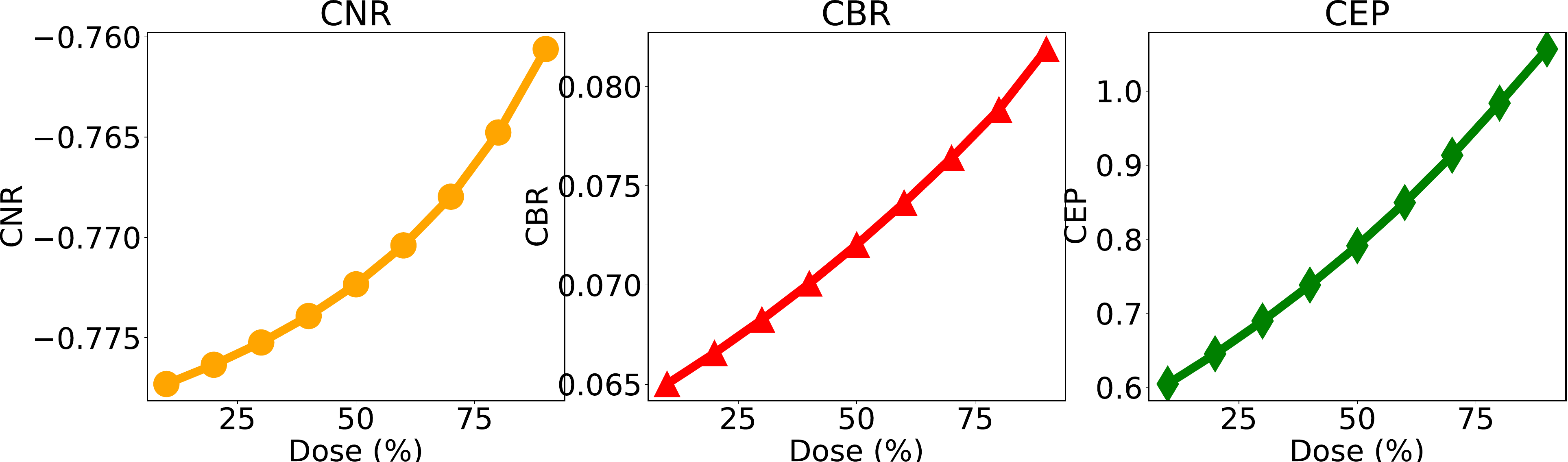}
\caption{Contrast uptake related quantitative metrics}
\label{contrast}
\end{figure}

\begin{figure}
\centering
\includegraphics[width=0.8\textwidth]{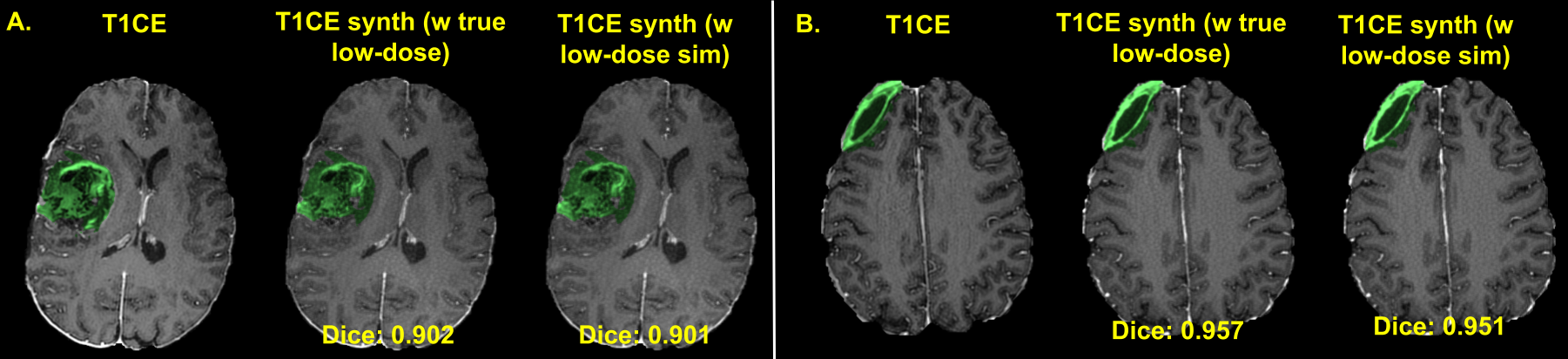}
\caption{Tumor segmentation (green overlay) on synthesized T1CE using real \& simulated low-dose in comparison to the tumor segmentation on ground truth T1CE. The corresponding Dice scores are also shown in the bottom.}
\label{tumorseg}
\end{figure}

\textbf{2) Tumor segmentation} Using the T1CE volumes synthesized in the above step, we perform tumor segmentation using the winning solution of BraTS 2018 challenge \cite{myronenko20193d}. Let $\mathbf{M}_{true}$, $\mathbf{M}_{ldose}$ and $\mathbf{M}_{ldose-sim}$ be the whole tumor (WT) masks generated using T1CE, T1CE-real-ldose and T1CE-synth-ldose (+ T1, T2 and FLAIR images) respectively. The mean Dice scores $Dice(\mathbf{M}_{true}, \mathbf{M}_{ldose})$ and $Dice(\mathbf{M}_{true}, \mathbf{M}_{ldose-sim})$ on the test set were $0.889 \pm 0.099$ and $0.876 \pm 0.092$ respectively. Fig. \ref{tumorseg} shows visual examples of tumor segmentation performance. This shows that the clinical utility provided by the synthesized low-dose is similar\footnote{p < 0.0001 (Wilcoxon signed rank test)} to that of the actual low-dose image.

\section{Discussions and Conclusion}
We have proposed a Gformer-based iterative model to simulate low-dose CE images. Extensive experiments and downstream task performance have verified the efficacy and clinical performance of the proposed model compared to other state-of-the art methods. In the future, further reader studies are required to assess the diagnostic equivalence of the simulated low-dose images. The model can be guided using physics-based models \cite{morkenborg2003quantitative} that estimate contrast enhancement level using signal intensity. This simulation technique can easily be extended to other anatomies and contrast agents.
\end{alphafootnotes}
%
%
\bibliographystyle{ieeetr}

\bibliography{main_arxiv}

\end{document}